%%%%%%%%%%%%%%%%%%%%%%%%%%%%%%%%%%%%%%%%%%%%%%%%%%%%%%%%%%%%
%%%%%%%%%%%%%%%%%%%%%%%%%%%%%%%%%%%%%%%%%%%%%%%%%%%%%%%%%%%%
%%%  On the Work of Henry P. Stapp.
%%%  quant-ph/0311158
%%%  November 2003
%%%
%%%
%%%   Plain TeX, 13 pages
%%%
%%%   Matthew J. Donald
%%%
%%%   web site:  http://www.poco.phy.cam.ac.uk/~mjd1014
%%%
%%%   e-mail  :  matthew.donald@phy.cam.ac.uk
%%%%%%%%%%%%%%%%%%%%%%%%%%%%%%%%%%%%%%%%%%%%%%%%%%%%%%%%%%%%
%%%%%%%%%%%%%%%%%%%%%%%%%%%%%%%%%%%%%%%%%%%%%%%%%%%%%%%%%%%%

\magnification=1200
\hsize=13cm
\def\newline{\hfil\break}  
\def\proclaim#1#2{\medskip\noindent{\bf #1}\quad \begingroup #2}
\def\endproclaim{\endgroup\medskip}
\def\tr{{\rm tr}}

\abovedisplayskip=3pt plus 1pt minus 1pt
\belowdisplayskip=3pt plus 1pt minus 1pt

\font\brm=cmbx12  \def\tilbf{\lower 1.1 ex\hbox{\brm \char'176}}

\font\trm=cmr12  
\def\tilrm{\lower 1.1 ex\hbox{\trm \char'176}}

\topskip10pt plus40pt
\clubpenalty=30

\headline={\hfil}
\footline={\hss\tenrm\folio\hss}

{\bf
\centerline{\ \  On the Work of Henry P. Stapp.
\ \footnote{*}{\tenrm quant-ph/0311158, November 2003.}}
\medskip

\centerline{Matthew J. Donald}
\medskip

\centerline{The Cavendish Laboratory,  Madingley Road,}  

\centerline{Cambridge CB3 0HE,  Great Britain.}
\smallskip

\centerline{ e-mail: \quad matthew.donald@phy.cam.ac.uk}
\smallskip
{\bf \hfill web site:\quad  
{\catcode`\~=12 \catcode`\q=9
http://www.poco.phy.cam.ac.uk/q~mjd1014
}\hfill }}
\medskip

For many years, Henry Stapp and I have been working separately and
independently on mind-centered interpretations of quantum theory.  In
this review, I discuss his work and contrast it with my own.  There is
much that we agree on, both in the broad problems we have addressed and
in some of the specific details of our analyses of neural physics, but
ultimately we disagree fundamently in our views on mind, matter, and
quantum mechanics.  In particular, I discuss our contrasting opinions
about the nature and randomness of quantum events, about relativity
theory, and about the many-minds idea.  I also suggest that Stapp's
theories are inadequately developed.

The theoretical analysis of the idea that there might be a link between
quantum mechanics and consciousness begins with the famous book by von
Neumann (1932).  von Neumann proposed a mathematical formalism according
to which there were two distinct ways in which a quantum system could
change with time.  On the one hand, there were abrupt indeterministic
``quantum events'' due to ``measurement'', in which, with appropriate
probability, one of the possible results of the measurement appears.  On
the other hand, for isolated and unmeasured systems, there was just the
continuous deterministic change described by the Schr\"odinger equation. 
This proposed duality in dynamics immediately raises the question of what
distinguishes ``measurements'' from other physical processes.  von
Neumann argued that measurement involved a chain of physical processes in
which the physical system to be measured interacts with a physical
measuring device, which ultimately interacts with light, which in turn
interacts with the eye of an observer.  Eventually information is carried
to the brain of the observer.  This suggests the possibility that it is
only when that information becomes conscious that something has occurred
which can make ``measurement'' different in kind from any other physical
process.

Stapp and I have both attempted to investigate this possibility by
proposing theoretical analyses to make sense of our conscious observations
of the world in the light of the evidence for quantum theory.  We have
both tried to understand the duality of quantum events and the
Schr\"odinger equation.  We both agree that brain processes are
fundamentally decoherent and that brain states should be analysed as
mixtures of essentially classical states (Donald 1990, 1992, Stapp 2000a,
2000b).  We have both noted the extent to which brains are unstable and
unpredictable dynamical systems (Donald 1990, 2002, Stapp 1993 chapter
6, 1999 Appendix A).  Both of us argue that neural uncertainties are
ultimately ``quantum'' in origin.  This leads both of us to stress the
multiplicity of quantum events which must be ``decided'', or of
``choices'' or ``measurements'' which must be ``made'', at every moment
in normal brain functioning.  

Beyond these basic agreements, there are however very significant
differences in our proposals.  According to my analysis, the abrupt
indeterministic quantum events are not physical occurences in a
conventional observer-independent sense.  They correspond rather to the
steps in the development of the mental structures of individual
observers.  My aim has been to provide an explicit abstract
characterization of the structure of an observer and to argue that human
beings can be described as possessing such structures.  These structures
are based on the idea of a temporal pattern of elementary abstract
quantum events which I call ``switchings'' (``determinations'' in Donald
1999).  I propose that there are no other quantum events.  Thus, in my
proposal, the problem of the characterization of consciousness and the
problem of the interpretation of quantum theory are both solved
together.  Stapp (1993, \S 6.7.4), by contrast, proposes that quantum
events are physical and occur in inanimate objects as well as in the
brains of observers.  This means that he encumbers himself with the
necessity of providing a characterization of the inanimate occurences of
the events if he is to complete his interpretation.  Were he to succeed
in doing this, he would already have managed to solve what is often seen
as the central problem in the interpretation of quantum theory, and his
comments on consciousness would seem to many (although not to me) to be
extraneous.

Stapp and I also disagree at a fundamental level about the randomness of
quantum events.  In Stapp (1993, \S 7.6), for example, he writes that,
``it is an absurdity to believe that the quantum choices can appear
simply randomly `out of the blue', on the basis of absolutely nothing at
all.''  Presumably because he believes that reality is not absurd, Stapp
uses this first claim to argue that consciousness intervenes in quantum
events to influence outcomes.  This second claim is bold, but in
Stapp (2001a), he moves toward the even bolder third claim, which he
attributes to both Copenhagen and von Neumann quantum theory, that ``the
choice of which question will be put to nature, is not controlled by any
rules that are known or understood within contemporary physics''.  Again,
in Stapp (2000b), he writes ``there is one element that [is] not governed
by any known law of physics, namely the choices to consent or not consent
at time
$t$ to putting to nature the question associated with the possible
experience $E(t)$''.

As far as the first claim is concerned, it is hard to avoid the idea that
in physics ultimately everything just does appear ``out of the blue''.  In
classical mechanics, for example, physics does not explain the initial
conditions, but merely provides the laws by which those initial conditions
develop.  One of the attractions of a many-minds approach, in my opinion,
is that it allows us to avoid the ``absurdity'' of requiring the entire
observed future to be encoded into the initial state of the universe. 
Instead, as I discuss in section 9 of Donald (1999), it is possible to
suppose that the initial state of the universe is a simple state --
perhaps even a vacuum state in the ultimate theory of everything -- and
to suppose that all the remaining information which constructs our
apparent individual reality is individually-observed information.  This
information is determined in our personal observations of the
genuinely-random, ``out of the blue'', outcomes of the quantum events
which make up our mental structures.

Stapp's second and third claims would seem to suppose that consciousness
is an extra-physical actor; directing physical dynamics by making
choices.  I find Stapp's position on this issue rather confusing.  In
section 6.7.2 of Stapp (1993), he rejects the somewhat similiar proposals
of Eccles (1986) as being explicitly dualistic and introducing a
``homunculus''.  The proposals of Walker (2000) are also similar, and I
have criticized them on the same grounds in Donald (2001b).  So Stapp and
I apparently agree that it is essential to avoid the idea of a little man
inside the brain directing the brain's thinking.  However the lesson of
modern neurophysiology seems to me to be that everything we experience is
directly reflected in the functioning and structure of our nervous
system.  Here ``everything we experience'' includes not only our thoughts
and feelings, but also our thoughts about our thoughts.  According to
this neurophysiological hypothesis, every human thought and action,
including choices, decisions, and self-analysis, can be explained in
terms of the functioning of the evolved physical brain.  One day, I
choose to take dessert because the firing from my appetite centers
dominates the firing from my prefrontal cortex and I say, ``I'll have the
chocolate mousse, please'', rather than, ``Nothing for me, thank you''. 
Another day, I will myself to abstain, because I have been sufficiently
disturbed by not being able to get my old trousers to fasten, that
thoughts of consequences outweigh thoughts of pleasure.  The ability of a
brain to talk about and apparently to decide its own behaviour is not
paradoxical, because it is limited, and because of the parallel and
modular structure of neural processing.  It is easy to understand why
such an ability should have evolved because it allows efficient analysis,
planning, and communication.

Stapp and I seem to agree that the evolved physical brain, as studied by
neurophysiologists, is not a deterministic machine.  Stapp proposes that
aspects of the random events which affect the path of the machine can be
chosen.  Yet, in Stapp (2000b), he writes, ``I do not intend to speculate
at this point about how the evaluation that lies behind this choice is
carried out.  At the present early stage in the development of the
science of the mind-brain system that question remains a project for
future research.''  If he is not simply invoking an extra-physical
homunculus at this point, however then at least we can ask whether
choices are supposed to be made using conventional neural circuitry --
in which case, it is certainly not clear to me where the requisite
circuitry is supposed to be -- or whether some other type of physical
process is supposed to be involved.

The events which initiate randomness in the brain include quantum
spreading in the paths of calcium ions (Stapp 1999).  Clearly there is no
neural circuitry which is capable of analysing those paths directly. 
Instead, Stapp's suggestion, at least in his earlier papers, seems to be
that the brain state becomes a superposition or mixture of different
developing possibilities until consciousness is reached, a choice occurs,
and the possibilities are reduced to a single outcome.  The trouble with
this idea is that the state becomes a mixture of different neural firing
patterns rather than a neural firing pattern analysing a mixture.  The
self-analysis of a such a mixture would require not just an entirely new
type of neurophysiology, but an entirely new type of physics.

\proclaim{example}{}  Suppose, as a result of quantum spreading, that in
one branch of the total quantum state of a brain the initiation of a
single neural firing is delayed, compared to another branch, by a mere
$10^{-5}$s relative to some other neural firing.  Because of the
metastability of neural firing, even a delay as short as this, on the
millisecond firing timescale, could be long enough to trigger radical
changes in the ultimate outcome of the developing neural response.  Yet
$10^{-5}$s is macroscopic on the picosecond timescale of molecular
vibrations in the warm wet brain.  This means that just this initial
difference is sufficient to make these branches of the total state
almost instantly become mutually decoherent.  They can then be
assigned their own separate density-matrix quantum states, which we shall
denote by $\rho_1$ and $\rho_2$.   So, in this situation, the relevant
part of the total state, by the time the firing patterns have developed
sufficiently for a choice between the outcomes to be possible in
conventional terms, will be a mixture of approximate form
$\rho = p_1 \rho_1 + p_2 \rho_2$ where $p_1$ and $p_2$ are the
conventional probabilities of the two possibilities.  The notation here
supresses important dependencies.  In non-relativistic quantum mechanics,
which is sufficient for our present purposes, $\rho$ depends on time and
on a choice of the space of operators considered relevant to the
situation, while in relativistic quantum field theory the time will
itself be specified by the space of operators.  However, at any time on
the conventionally-recognised timescale of conscious processing and for
any space of operators localized within the brain of the individual
considered, $\rho$ will have a decoherent decomposition of the suggested
form.  It follows that no conventional physical process within the brain
will be able to cause a (generalized) ``collapse'' from $\rho$ to either
$\rho_1$ or $\rho_2$.  Because of decoherence, $\rho_1$ and $\rho_2$ have
become dynamically independent.  The splitting into decoherent states is
a locally irreversible process due to the dissipation of information. 
\endproclaim  

In his earlier papers (1982, 1993 chapter 6, 1995), Stapp seems to suggest
that somehow the structure of $\rho$, or possibly of $\rho_1$ or $\rho_2$,
directs the collapse.  How this is supposed to work is not explained.  In
his later papers (1999, 2000a, 2001a), Stapp suggests the quantum Zeno
process as the mechanism.  This suggestion has been picked up by Jeffrey
M. Schwartz (Schwartz and Begley 2002).  In his treatment of
obsessive-compulsive disorder patients, Schwartz appeals to Stapp's work,
along with the Buddhist concept of mindfulness, to argue for the reality
of free will.  Nothing I say here should be taken as a criticism of
Schwartz's plausible and apparently-successful treatment, which involves
patients learning to challenge their intrusive thoughts.  Nor do I
criticize the wealth of interesting evidence reviewed by Schwartz and
Begley about the possibility, throughout life, of quite large-scale
changes in neural connectivity.  This evidence allows them to justify a
strong form of the conclusion that, ``it is the life we lead that creates
the brain we have''.  I only disagree with the framework which Schwartz
uses to explain the meaning of challenging one's own thoughts.

Johnjoe McFadden (2000) has independently invoked the quantum Zeno effect
in an attempt to solve problems in biological science.  In Donald
(2001a), I criticise McFadden's work on the grounds that, in as far as the
Zeno effect (or ``inverse Zeno effect'' to use McFadden's language) is a
physical effect, it can occur only in very carefully established
circumstances, precisely set up in order to achieve a given end.  Suppose,
in the notation of the example above, that $\rho_2$ represents the outcome
which is to be chosen.  Then, as I explain in Donald (2001a), the Zeno
effect does provide a dynamics formed using a sequence of carefully chosen
projections, corresponding to von Neumann's quantum events, which, with
probability one, will drive the total state towards the state $\rho_2$. 
That dynamics, however, is not the well-understood biologically-evolved
dynamics produced by the interactions of the ions and atoms and molecules
and electric fields of the human brain.  It is a purely theoretical
dynamics which depends on working towards the desired outcome right from
the start of the initial quantum spreading.  How and when the choice is
supposed to result in the construction and action of the projection
operators still has to be explained.  The projections need to be defined
to the precision of an individual wavefunction.  Above atomic scales,
no biological mechanism can control or even repeat states at this level of
precision. 

If we want to observe the quantum Zeno effect in the laboratory, then we
need to build an apparatus (e.g.~Itano et al.\ 1990).  Using von Neumann's
idea of ``measurement'' as a chain of physical processes, it would seem
that such a physical apparatus will itself have a dynamics which can be
understood in terms of the conventional laws of physics described by the
Schr\"odinger equation, and which can be used to explain the observed
effects.  Models of this type are discussed in chapter 8 of Namiki,
Pascazio, and Nakazato (1997), in section 3.3.1 of  Giulini et
al.~(1996), and in Gurvitz (2002). According to this picture, Stapp needs
to tell us what apparatus biology has built into our brains which allows
us to use the quantum Zeno effect to make choices.  Otherwise, he would
seem to be supposing that brains are somehow physically special and that,
for some reason, they cannot themselves be analysed by external
observers.  He would have invented a homunculus with access to a space of
projection operators defined on an atomic scale.  How is that homunculus
supposed to choose which projection he wants to employ next?  How does
he control individual wavefunctions or the projections which correspond
to them?  

von Neumann's abrupt changes seem to require the choice of a question;
the choice of what is to be measured.  At the heart of Stapp's work is the
idea that making that choice is what consciousness does.  However, I
believe that when we make a choice we are doing mental work involving
ordinary physical neural processing; just as when we express a question
in words we use the linguistic mechanisms which are available in our
brains as a result of the life we have led.  In my opinion, despite the
interesting ideas of William James (Stapp 1993) and Harold Pashler (Stapp
2001a), neurophysiology is more fundamental than psychology.  My basic
objection to Stapp's work is that he does not appear to have made any
connection between the representation of choices by physical neural
processing -- or indeed any other cellular process -- and their
representation as projection operators to be measured.  Only for an
extra-physical homunculus, does the consideration of a choice not involve
physical operations.  And however the choices are made, the ultimate
purpose of choosing must lie in the physical consequences of the choice;
in other words in the change in the global quantum state.  Stapp has not
explained how he supposes such changes are limited.  Why should they be
restricted to changes within a brain?  If mental forces can effectively
decide the trajectories of atoms or molecules inside a brain, why can
they not decide the trajectories of electrons in a laboratory or of prey
in the ocean?  What determined the point in evolutionary history when
brains are supposed to have started to be able to make choices?  And
finally, even if we do suppose that mind is some sort of extra-physical
homunculus, we have only managed to introduce a new mystery.  If there is
a problem of free will, then it will still be there if we can ever get to
the point of analysing the operations of the homunculus.

As I see it, the central problem of the interpretation of quantum theory
is to explain and characterize the existence, or apparent existence, of
``quantum events'', or, in other words, of the process referred to as
state ``collapse'', or ``wave-packet reduction'', or ``von Neumann's
process 1''.  How is it, for example, that an electron, despite going
through a double slit as an extended wave, always appears to make a
well-localized impact on a screen at only one of many possible places, so
that the extended wave apparently ``collapses'' to a localized state?  A
characterization of collapse should tell us what possibilities arise --
this is ``the preferred basis problem''.  The characterization should be
well-defined and unambiguous.  It should be explained how one collapse
leads on to the next and the probability of a given collapse should also
be well-defined.  With such a theory, it will no longer be possible to
invoke the quantum Zeno effect as a clever trick by which arbitrary
sequences of collapses can be used to attain any desired outcome. 
Instead, there will be specific circumstances in which the effect, or the
appearance of the effect, will arise as a consequence of the specific
collapses which actually occur, or which appear to occur.

Just as one of the most fundamental questions in the philosophy of mind is
whether mental events are merely how neurophysiological events appear, so
one of the most fundamental questions in the philosophy of quantum theory
is whether von Neumann's indeterministic events are merely how the
continuous changes appear in specific circumstances.  My starting point
is to answer both questions affirmatively, and therefore I have tried to
develop a theory characterizing ``appearance''.  Stapp certainly answers
the second negatively and therefore owes us an analysis of quantum
events.  He is ambiguous about the first.  In Stapp (1993 \S 7.5), he
claims that there is, ``An isomorphic connection [\ ] between the
structural forms of conscious thoughts, as described by psychologists, and
corresponding actualized structural forms in the neurological patterns of
brain activity, as suggested by brain scientists.''  On the other hand,
on the same page, he claims to have provided for, ``A mechanical
explanation of the efficacy of conscious thoughts''.  I have no idea to
what this is supposed to refer.

In Stapp (1993 \S 1.10), Stapp states that his theory ``makes
consciousness causally effective, yet it is fully compatible with all
known laws of physics, including the law of conservation of energy.'' 
Stapp does not justify this statement.  In general, energy is not
conserved in individual quantum jumps.  Average total energy may be
conserved if the projections involved commute with the global
Hamiltonian.  Leaving aside the commutation question, however, this
would require that ``causal effectiveness'' produces the same averages as
conventional quantum probabilities.  In Stapp (1995), Stapp admits that,
``No attempt is made here to show that the quantum statistical laws will
hold for the aspects of the brain's internal dynamics controlled by
conscious thoughts''.  While it may be appropriate, as he suggests in the
same paper, to make an assumption of ignorance about the external causes
of external events, it would be absurd to suppose that conscious thoughts
can be efficacious but without external consequences.  The suggestion
mentioned above that the initial state of the universe might be a vacuum
state is certainly not compatible with the idea that energy is conserved
(or appears to be conserved) in individual quantum events (or observed
events).

Stapp suggests that state collapse is driven by conscious choice, but does
not explain what conscious choice involves.  Ulrich Mohrhoff (2001)
criticizes Stapp's work and is replied to by Stapp (2001b).  In my
opinion, Mohrhoff's proposals are at least as weak as Stapp's. 
Mohrhoff agrees with Stapp that, ``the choice of which question will be
put to  nature \dots is not governed by the physical laws of contemporary
physics'', but he bases his analysis of quantum events on the idea of the
existence of definite ``facts'' and he explicitly denies that ``facts''
can be characterized.  

In my own work, I have attempted to give an {\it abstract\/}
characterization of ``facts''.  Thus, instead of trying to define a
particular set of projection operators giving rise to the observed
collapses, I have defined an abstract pattern of projection operators
which is capable of expressing the observed information.  Such a
pattern can be constituted by many possible sets of projections, and the
entire set of these sets for a given pattern is central to the
definitions I provide.  By working at a high level of abstraction in this
way, it is possible to avoid much arbitrariness in the definitions.  I
propose that each of us exists as an individual developing pattern of
information.  The development is stochastic, with probabilities defined
by quantum mechanical laws.  According to these laws, the experience of
each individual observer is the experience of observing a particular,
identified, discrete stochastic process.

While I work at a high level of abstraction, I work with simple
elements.  In other words, I have built up patterns of information using
projections considered as yes-no questions.  This is speculation.  It is
just a hypothesis that information in the brain is constituted as a
pattern of yes-no questions.  Making a specific hypothesis, however, does
allow specific technical questions to be addressed and allows a theory
to be developed and its defects to be revealed.

My theory is a many-minds theory.  This means that each of us has our own
pattern of observed ``quantum events''.  Consistency between
mutually-aware observers is a consequence of the nature of quantum
probability.  My theory is dualistic in the sense that there are physical
laws and there are observers, but there are no mental computations without
observable physical structure.  My theory is epiphenomenalistic in the
sense that a mind does not direct a pattern of observed physical events,
rather it has to make sense of such a pattern as it unfolds.  Ultimately,
however, my theory should probably be considered as idealistic because, in
its final form, the central structures in the theory are mental
structures.  Physics just supplies the probabilities by which those
mental structures change.  Mental structures give meaning to their
realities by understanding themselves in terms of observable physical
structures and observed physical events.  I propose that, with an
appropriate definition of mental structure, the nature of quantum
probability will make it likely that if awareness is attained, it will
understand itself in such terms.  The centrality of mind allows the
often-asked question of how an epiphenomenal consciousness could have
evolved, to be side-stepped. Instead, I suggest that we should look for
the most likely way in which a structure sufficiently complex to be
self-aware could appear to have arisen under simple natural laws.

Relativity is another major issue which has led Henry Stapp and I to take
very different paths.  With the extension to relativistic quantum field
theory, the continuous deterministic unitary change at the heart of
quantum theory does become manifestly compatible with the theory of
special relativity.  Abrupt events however are much more problematic. 
Special relativity requires that no change can be communicated at a speed
faster than that of light.  Yet von Neumann's abrupt events apparently
happen instantaneously across the entire universe.  Moreover, there seems
to be considerable empirical evidence for instantaneous non-local changes
in a rather peculiar form of correlation information about what might
happen under various independent choices.  Stapp has generated much
debate (e.g.~Unruh 1997, Mermin 1997) by arguing that quantum mechanics
is therefore non-local (Stapp 1993 chapter 1, 1997, 2000c, 2002). 
Interesting as the subleties of this debate may be, the most direct way
to resolve the problem that it addresses is to assume that the symmetry
of special relativity is broken by some objective sequence of
hypersurfaces of simultaneity.  This is the assumption that Stapp (1993
\S 4.5.13, 1997, 2001a) makes.  

Citing Tomonaga (1946) and Schwinger (1951), Stapp (2001a) claims that
this assumption ``does not disrupt the covariance properties of the {\sl
empirical predictions\/} of the theory'' [his italics].  The precise
relevance of these old papers, the technical validity of which has been
called into question by Torre and Varadarajan (1998), is not entirely
clear to me.  The arguments of Tomonaga and Schwinger are concerned with
generalizing the Schr\"odinger equation to allow for arbitrary spacelike
hypersurfaces, rather than with an analysis of individual quantum events. 
A more modern approach to relativistic quantum field theory uses the
language of local algebraic quantum field theory (Haag 1992).  According
to this theory, empirical predictions made in two spacelike-separated
regions $\Lambda_1$ and $\Lambda_2$ can be expressed by commuting
projection operators $P_1$ and $P_2$.  Then, if $\rho$ is the initial
quantum state, the probability of the result corresponding to $P_1$ being
seen in $\Lambda_1$ and the result $P_2$ being seen in $\Lambda_2$ is
given by $\rho(P_1 P_2)$ in the notation used by mathematicians. 
(Physicists would think of $\rho$ as a density matrix and would write
$\tr(\rho P_1 P_2)$.)

In this context, following the observation in $\Lambda_1$, von Neumann's
abrupt event would correspond to the replacement of $\rho$ by $\rho_1 =
P_1 \rho P_1/ \rho(P_1)$ and then, if the observation in $\Lambda_2$ is
considered to be subsequent, $\rho_1$ would be
replaced by $\rho_{12} = P_2 \rho_1 P_2/ \rho_1(P_2) = P_2 P_1 \rho P_1
P_2/\rho(P_1 P_2)$.  As $P_1$ and $P_2$ commute, this is symmetric under
interchange of $1$ and $2$, and so, consistent with Stapp's claim, does
not depend on the ordering of spacelike separated events.

Apart from the question, raised above, of whether the probabilities of
events in Stapp's theory are equal to the conventional quantum
probabilities given by state expectation values, there are two
problems with Stapp's reliance on this argument.  The first is with the
assumption that all quantum events can be divided into classes associated
with some foliation of spacetime into spacelike hypersurfaces.  In
relativistic quantum field theory, events are associated with spacetime
regions rather than with spacetime points.  If there are too many events
too close together, then there may be no simple way of ordering the
corresponding regions.  This is why, in my work (Donald 1995), I use the
full causal structure of the relations between spacetime sets in which
events occur.  

The second problem is with the identification of the ``initial quantum
state'' -- the state $\rho$. 

Suppose that I try to model my observations of some particular local
macroscopic system by associating a quantum state $\sigma_1$ with that
system.  $\sigma_1$ will depend on my observations and my knowledge of the
system.  This means that $\sigma_1$ will already be a ``collapsed''
state, like $\rho_1$, rather than ``uncollapsed'', like $\rho$.  Indeed,
the only truly ``uncollapsed'' state would be the actual initial state of
the universe, which is the state Everett refers to as the universal
wavefunction.  $P_2 P_1 \rho P_1 P_2/\rho(P_1 P_2)$ may be symmetric
under interchange of $1$ and $2$, but it does still depend on $P_1$. If
Stapp's suggestions are correct, $\sigma_1$ will depend on the time in the
universal clock defined by Stapp's objective sequence of hypersurfaces of
simultaneity and on non-local events at arbitrary distances; involving
perhaps, conscious observers in other galaxies.

Without a complete theory describing the nature of quantum events, it is
difficult to decide on the significance of this dependence of local
observed states on the supposed external events.  Relativistic quantum
field suggests that correlations between distant events are ever-present
(Clifton and Halvorson 2000).  Many of these correlations may be
individually negligible, but if instantaneous action at a distance is
allowed, then, at any moment, the effect of infinitely many distant
events may need to be taken into account.  In particular, this might make
it surprising that the passage of time on Stapp's universal clock, if it
existed, should not have been observed through some sort of
frame-dependent effect on system dynamics.

An alternative way of looking at this issue would be to suppose that in
some region $\Lambda_1$ (the ``external'' region) there is a range of
possible results, corresponding to projections $(P_1^n)_{n=1}^N$.  As
long as $\sum_{n=1}^N P_1^n = 1$, result $P^n_1$ will have probability
$\rho(P^n_1)$ in a global state $\rho$, and the expected state in a
spacelike-separated region $\Lambda_2$ will be $\sum_{n=1}^N  \rho(P^n_1)
P^n_1 \rho P^n_1/\rho(P^n_1) = \sum_{n=1}^N  P^n_1 \rho P^n_1$.  As
$P_1^n$ commutes with every operator local to $\Lambda_2$, this expected
state is equal to $\rho$ on $\Lambda_2$.  This indicates that external
events have no average effect in $\Lambda_2$.  However, going from this
argument to the conclusion that even infinitely many individual choices
of projections like $P^n_1$ acting on the state $\rho$ will have no
observable effect within $\Lambda_2$ seems rather like imagining that it
would help a seasick man to be told that the average height of the sea is
constant.

It is not straightforward to reconcile the apparent empirical evidence for
instantaneous non-local changes in correlation information with the
apparent evidence for the Lorentz invariance of physical processes.  My
own approach takes as fundamental the information possessed by individual
localized observers.  Each observer assigns, to any system, the expected
quantum state given the information s/he currently possesses.  In
general, these states are not compatible.  Although Alice may know that
her distant colleague Bob will have performed an experiment and will have
found a result, she can only assign to the system which he is observing a
state which expresses her ignorance until such time as she learns what
experiment he chose to perform and what results he found.  This has
suggested to many (e.g. Wolfe 1936, Wigner 1961, Peierls 1991, Fuchs
2002) that quantum states are merely states of knowledge.  Unfortunately,
such a position becomes problematic when we try to understand
psycho-physical parallelism (Donald 2002).

Quantum states seem to play two roles; on the one hand describing how
things are, and on the other describing what we know about how things are
and about how they might be.  Alice and Bob assign different states to
each other in ways which suggest that, for Alice, Bob's existence has to
be as indefinite as her knowledge of his observations.  This provides a
primary motivation for a many-minds approach.  If individual local
observers are considered separately, then we can suppose that for Alice,
all of Bob's possibilities will continue to exist until she is in a
position to see his results for herself.  By this means, the problem of
compatibility between special relativity and quantum theory can be
reduced from dealing with events which are supposed to occur
instantaneously across the entire universe, to dealing with events which
occur within the structure of individual observers (cf.~Tipler 2000,
Timpson and Brown 2002).  Of course, no single observer should be singled
out.  If Alice believes that all of Bob's alternative possibilities
continued to exist after the instant of Bob's observation, then she should
also be prepared to allow that all of her own alternative possibilities
will also have continued to exist.

Stapp has always rejected many-worlds interpretations.  He claims that
such interpretations have many technical problems to overcome (Stapp
1993, \S 1.13).  In Stapp (2001c), for example, he reviews some of the
well-known problems with the simplistic idea of a many-worlds theory
which depends on a specific preferred orthonormal basis.  It was precisely
in order to avoid the kind of problem which he discusses in that paper
that I developed a theory in terms of abstract patterns of information
expressed not by a precise choice of wavefunction basis, but by ranges of
properties of density matrices (Donald 1986, 1990).  

Stapp also raises metaphysical problems with many-worlds interpretations;
in particular problems concerning the nature of probabilities.  For
example, in Stapp (1999) he writes, ``In the evolving wave
function of Everett the various branches do evolve independently, and
hence might naturally be imagined to have different `minds' associated
with them, as Everett suggests.  But these branches, and the minds that
are imagined to be properties of these branches, are all simultaneously
present.  Hence there is no way to give meaning to the notion that one
mind is far more likely to be present at some finite time than the
others.''. 

This is a much discussed problem (Loewer 1996, Vaidman 1996, Saunders
1998). My response is to emphasize the primacy of the experience of
probability.   Life is a game of chance. If we can give any meaning to
probability at all, then we can give meaning to the notion that we are
far more likely to observe one possible future event than another.  In
other words, if, as Stapp and I both propose, we accept that
consciousness is a significant aspect of reality, then we should be
prepared to examine the idea of ``simultaneous presence''.  Although, of
course, there is a sense in which other branches are present, only my
current branch is present to me.  Probability is something which,
according to my theory, is experienced in individual branches.  The
experience of an individual branch is genuinely unpredictable.

I would similarly emphasize the importance of the experience of free
will.  Schwar\-tz and Begley (2002) suggest that free will is incompatible
with determinism, but whe\-ther all your actions have already been
forseen by some omniscient deity, or whether your brain is abuzz with
quantum jumpings has no relevance to your apparent ability to decide
whether to call heads or tails when a coin is tossed.  Until you call,
you know you are free to change your mind.

In my opinion, the lesson of Zeno's original classical paradoxes is
that it is a mistake to attempt to use metaphysics to foreclose options in
theoretical physics.  Stapp and I agree that normal neural functioning
provides a stream of quantum events.  In my view, the die roll for each
of us as each of our quantum events occurs.  My primary goal has been to
understand the technical problems of how to define such events (Donald
1990) and patterns of them (Donald 1995), and of how to define
probabilities for these events in a way compatible with the mathematics
of quantum theory and the assumption that our observations are typical
(Donald 1986, 1992, 1999).  If that task has been successfully completed,
then it is time to turn metaphysician and investigate what might be
implied about the nature of reality.  Of course, the implications may be
strange or even disturbing, and may give us reason to prefer an
alternative theory if we have one, but I do not believe that metaphysics
by itself can provide a convincing refutation.  Refutation, for the sort
of theory I am proposing, would, I believe, require flaws in the
technical details, as discussed in section 1 of Donald (1999), or
empirical evidence that quantum theory does not apply at the macroscopic
level.  There may be many other reasons, such as the speculative nature
of the theory and its complexity, to prefer an alternative theory, but it
is hard to know how seriously these reasons should be taken if the
alternative has not reached a similar level of development.  It is
precisely because of the importance and difficulty of constructing a
complete theoretical structure that it is so disappointing that Stapp
fails to provide more than sketches of his ideas about the mathematical
structure of thought (Stapp 1993 Appendix), about selection of top-level
codes (Stapp 1982) and selection processes (Stapp 1995), about the role
of the electromagnetic field (Stapp 1999), and about the quantum Zeno
effect (Stapp 1999, 2000a, 2001a).

\

\

\proclaim{References}{}
\endproclaim

\frenchspacing
\parindent=0pt

{\everypar={\hangindent=0.75cm \hangafter=1} 

Clifton, R. and Halvorson, H. (2000) ``Entanglement and open systems in
algebraic quantum field theory.'' {\sl  quant-ph/0001107}.  Printed: {\sl
Stud. Hist. Philos. Mod. Phys. \bf 32}, 1--31 (2001).

Donald, M.J. (1986)  ``On the relative entropy.''  {\sl Commun. Math. Phys.
}{\bf 105},  13--34.

Donald, M.J. (1990) ``Quantum theory and the brain.''  {\sl Proc.
R. Soc. Lond. \bf A 427},  43--93.

Donald, M.J. (1992) ``A priori probability and localized observers.'' {\sl 
Found. Phys. \bf 22}, 1111--1172.

Donald, M.J. (1995) ``A mathematical characterization of the physical
structure of observers.'' {\sl  Found. Phys. \bf 25}, 529--571.

Donald, M.J. (1999)  ``Progress in a many-minds interpretation of
quantum theory.'' {\sl  quant-ph/9904001}

Donald, M.J. (2001a)  ``A review of Johnjoe McFadden's book {\sl Quantum
Evolution}.'' {\sl  quant-ph/0101019}

Donald, M.J. (2001b)  ``A review of Evan Harris Walker's book {\sl The
Physics of Consciousness}.''  {\sl Psyche \bf 7}

{\hfill{ http://psyche.cs.monash.edu.au/v7/psyche-7-15-donald.html
}\hfill}

Donald, M.J. (2002)  ``Neural unpredictability, the interpretation of
quantum theory, and the mind-body problem.''  {\sl quant-ph/0208033}

{\hfill Papers and related material of mine are also available
on my web site\hfill}

{\hfill {\catcode`\~=12 \catcode`\q=9
http://www.poco.phy.cam.ac.uk/q~mjd1014 } \hfill}
\smallskip

Eccles, J.C. (1986)  ``Do mental events cause neural events analogously to
the  probability fields of quantum mechanics?''  {\sl Proc. R. Soc. Lond. \bf
B 227},  411--428.

Fuchs, C.A. (2002)  ``Quantum mechanics as quantum information (and
only a little more).''  {\sl quant-ph/0205039}

Giulini, D., Joos, E., Kiefer, C., Kupsch, J., Stamatescu, I.-O., and
Zeh, H.D. (1996) {\sl Decoherence and the Appearance of a Classical
World in Quantum Theory}  (Springer).

Gurvitz, S.A. (2002) ``Quantum description of classical apparatus: Zeno
effect and decoherence.'' {\sl  quant-ph/0212155}

Haag, R. (1992)  {\sl Local Quantum Physics.} (Springer)

Itano, W.M., Heinzen, D.J., Bollinger, J.J., and Wineland, D.J. (1990) 
``Quantum Zeno effect.''  {\sl Phys. Rev. \bf A 41}, 2295--2300.

Loewer, B. (1996) ``Comment on Lockwood.'' {\sl  Brit. J. Phil. Sci. \bf
47},  229--232.

McFadden, J. (2000) {\sl Quantum Evolution: Life in the Multiverse.} 
(HarperCollins)

Mermin, N.D. (1997) ``Nonlocal character of quantum theory?'' {\sl 
quant-ph/9711052}. Printed: {\sl Amer. J. Phys. \bf 66}, 920--924 (1998).

Mohrhoff, U. (2001) ``The world according to quantum mechanics (or, the
18 errors of Henry P. Stapp).'' {\sl  quant-ph/0105097}.  Printed: {\sl
Found. Phys. \bf 32}, 217--254 (2002).

Namiki, M., Pascazio, S., and Nakazato, H.  (1997)  {\sl  Decoherence
and Quantum Measurements} (World Scientific).

von Neumann, J. (1932)  {\sl Mathematische Grundlagen der
Quantenmechanik} (Springer) English translation (1955) {\sl
Mathematical Foundations of Quantum Mechanics} (Princeton).

Peierls, R. (1991) ``In defence of `measurement'.'' 
{\sl Physics World}, January 19--20.

Saunders, S. (1998) ``Time, quantum mechanics, and probability.''
{\sl  Synthese, \bf 114},  373--404.  {\sl quant-ph/0111047}

Schwartz, J.M. and Begley, S. (2002) {\sl The Mind and the Brain:
Neuroplasticity and the Power of Mental Force.} (HarperCollins)

Schwinger, J. (1951) ``The theory of  quantized  fields, I.''  {\sl
Phys. Rev. \bf 82}, 914--927.

Stapp, H.P. (1982) ``Mind, Matter, and Quantum Mechanics.'' {\sl 
Found. Phys. \bf 12}, 363--399.  Reprinted as chapter 4 of Stapp (1993).

Stapp, H.P. (1993) {\sl Mind, Matter, and Quantum Mechanics.} (Springer).
A new edition of this book has been announced.

Stapp, H.P. (1995) ``Chance, choice, and consciousness: The role of mind
in the quantum brain.'' {\sl  quant-ph/9511029}

Stapp, H.P. (1997) ``Nonlocal character of quantum theory.''  {\sl Amer. J.
Phys. \bf 65,} 300--304.

Stapp, H.P. (1999) ``Quantum ontologies and mind-matter synthesis.'' 
\newline {\sl  quant-ph/9905053}.  Printed in {\sl Quantum Future,} eds.
P. Blanchard and A. Jadczyk, (Springer 1999).

Stapp, H.P. (2000a) ``Decoherence, quantum Zeno effect, and the efficacy of
mental effort.'' {\sl quant-ph/0003065}

Stapp, H.P. (2000b) ``The importance of quantum decoherence in brain
processes.'' {\sl  quant-ph/0010029}

Stapp, H.P. (2000c) ``Bell's theorem without hidden variables.'' {\sl 
quant-ph/0010047}

Stapp, H.P. (2001a) ``Quantum theory and the role of mind in nature.'' 
\newline {\sl quant-ph/0103043}.  Printed:  {\sl Found.
Phys. \bf 31}, 1465--1499 (2001).

Stapp, H.P. (2001b) ``The 18-fold way.'' {\sl quant-ph/0108092}. 
Printed:  {\sl Found. Phys. \bf 32}, 255--266 (2002).

Stapp, H.P. (2001c) ``The basis problem in many-worlds theories.'' {\sl 
quant-ph/0110148}.  Printed: {\sl Can. J. Phys. \bf 80}, 1043--1052
(2002). 

Stapp, H.P. (2002) ``A Bell-type theorem without hidden variables.'' 
\newline {\sl quant-ph/0205096}.

{\hfill Papers and related material by Henry Stapp are also available
on his web site\hfill}

{\hfill {\catcode`\~=12 \catcode`\q=9
\hfill http://www-physics.lbl.gov/q~stapp/stappfiles.html } \hfill}
\smallskip

Timpson, C.G. and Brown, H.R. (2002)  ``Entanglement and relativity.''
\newline {\sl  quant-ph/0212140}. To appear in {\sl Understanding Physical
Knowledge,} eds V. Fano and R. Lupacchini.

Tipler, F.J. (2000)  ``Does quantum nonlocality exist?  Bell's theorem and
the many-worlds interpretation.''  {\sl  quant-ph/0003146}

Tomonaga, S. (1946)  ``On a relativistically invariant formulation of the
quantum theory of wave fields.''  {\sl Prog. Theor. Phys. \bf 1}, 27--42.

Torre, C.G. and Varadarajan M. (1998)  ``Functional evolution of free
quantum fields.'' {\sl hep-th/9811222}. Printed: {\sl Class. Quant.
Grav. \bf16,} 2651--2668 (1999).

Unruh, W.G. (1997)  ``Is quantum mechanics non-local?''  {\sl 
quant-ph/9710032}

Vaidman, L. (1996)  ``On schizophrenic experiences of the neutron.'' 
\newline {\sl  quant-ph/9609006}.  Printed:  {\sl Int. Stud. Phil. Sci.
\bf 12}, 245--261 (1998).

Walker, E.H. (2000) {\sl The Physics of Consciousness: The Quantum
Mind and the Meaning of Life.}  (Perseus)

Wigner, E.P. (1961)  ``Remarks on the mind-body question.''  In {\sl
The Scientist  Speculates,} ed. I.J. Good, 284--302. (Heinemann)

Wolfe, H.C. (1936) ``Quantum mechanics and physical reality.'' 
{\sl Phys. Rev. \bf 49}, 274. 

}

\end